\documentclass[preprint,nofootinbib,aps,showpacs,preprintnumbers]{revtex4-1}

\usepackage{graphicx}
\usepackage{dcolumn}
\usepackage{epsfig}
\usepackage{epstopdf}
\usepackage{amssymb}
\usepackage{amsfonts}
\usepackage{xcolor}
\usepackage{float}
\usepackage{hyperref}
\usepackage{physics}
\usepackage{comment}
\usepackage[caption=false]{subfig}
\usepackage{soul}
\usepackage[bottom]{footmisc}

\usepackage{lineno}
\usepackage{amsmath}
\usepackage{dirtytalk}

\hypersetup{
	colorlinks = true,
	linkcolor = blue,
	citecolor = blue,
	urlcolor = blue
}
\allowdisplaybreaks
\setstcolor{red}

\begin{document}

\title{Effect of light-assisted tunable interaction on the position response function of cold atoms}

\author{Anirban Misra$^1$}
\email{anirbanm@rrimail.rri.res.in}
\author{Urbashi Satpathi$^2$}
\email{urbashi.satpathi@gmail.com}
\author{Supurna Sinha$^1$}
\email{supurna@rri.res.in}
\author{Sanjukta Roy$^1$}
\email{sanjukta@rri.res.in}
\author{Saptarishi Chaudhuri$^1$}
\email{srishic@rri.res.in}

\affiliation{$^1$ Raman Research Institute, C. V. Raman Avenue, Sadashivanagar, Bangalore-560080, India.}
\affiliation{$^2$Department of Physics and Materials Science Engineering, Jaypee Institute of Information Technology, A-10, Sector 62, Noida, UP-201309, India.}

\date{\today}
\begin{abstract}
\noindent The position response of a particle subjected to a  perturbation is of general interest in physics. We study the modification of the position response function of  an ensemble of cold atoms in a magneto-optical trap in the presence of tunable light-assisted interactions. We subject the cold atoms to an intense laser light tuned near the photoassociation resonance and observe the position response of the atoms subjected to a sudden displacement. Surprisingly, we observe that the entire cold atomic cloud
undergoes collective oscillations. We use a generalised quantum Langevin approach to theoretically analyse the results of the experiments and find good agreement.
\end{abstract}

\maketitle

\section{Introduction}
\noindent Cold atoms are excellent candidates for precision measurements \cite{inguscio2013atomic} and for studying time dependent response \cite{Morawetz2014reponse, bhar2022measurements} of atoms to external perturbations. Tracking the position response function of cold atoms is expected to enhance the understanding of many-body physics with cold atoms and quantum gases.
One particular area of interest has been precision molecular spectroscopy using ultra-cold atoms which is known as photoassociation (PA) spectroscopy 
\cite{thorsheim1987laser,miller1993photoassociation,lett1995photoassociative,Stwalley_Wang_1999,weiner1999experiments,lett1993spectroscopy,jones2006ultracold,C7CP08480C}. The traditional detection scheme in such experiments has always been to measure a trap loss where the photoassociated molecules leave the trap \cite{thorsheim1987laser,lett1993spectroscopy,miller1993photoassociation,DalgarnoPhotoloss1998,Stwalley_Wang_1999,weiner1999experiments,Killian2021}. 
It is well known that near such PA transitions the inter-atomic interaction strength is significantly modified due to dipole-induced dipole interactions even when the bound state is not created \cite{van1998photoassociation,fatemi2000observation,2011_Ye,taie2016feshbach,2015_Julienne}.
\par
Here we report a technique for detecting changes in inter-atomic interactions near a photoassociation resonance, without relying on existing trap-loss measurements \cite{thorsheim1987laser,lett1993spectroscopy,miller1993photoassociation,DalgarnoPhotoloss1998,Stwalley_Wang_1999,weiner1999experiments,Killian2021}. Our technique involves observing the atomic position response to a sudden external light pulse in a Magneto-Optical Trap (MOT), offering an alternative approach to study light-assisted collisions between ultra-cold atoms. By adjusting the cooling laser parameters and the magnetic field gradient, we can control the damping and spring constants of the atomic motion. Such response measurements of cold atoms in the MOT have been the subject of interest of many past studies \cite{kim2005measurement,beilin2010diffusion, hodapp1995three, deng2007brownian} including some recent ones \cite{satpathi2017quantum,bhar2022measurements,MOON20171,baek2023measurement}. Generally, motion of the atoms in the MOT is well understood and the response function can be modelled efficiently using a generalized Langevin formalism \cite{Ford1988}. The corrections to such measurements taking into account the beyond two-level approximation has also been the subject of a recent work \cite{baek2023measurement}. 
Here we present an experimental study and the corresponding theoretical treatment of the motion of the atoms trapped in a MOT under the influence of an additional laser beam which tunes the interactions between the atoms. In the latter part of the paper, we present the details of our experimental sequence, the observations and the theoretical analysis.
\par
We notice a qualitatively interesting transition in the behaviour of position response function of the trapped cloud. In a parameter space which is overdamped, surprisingly the emergence of collective oscillations (which occurs in underdamped regime) is observed in the position response function as a result of tuning the interactions between the atoms. The cloud can be viewed as being in contact with a bath of optical molasses where we tune the inter-atomic interaction using the PA beam in the vicinity of a molecular resonance \cite{2015_Julienne}.

\section{Measurement of position response function}\label{section2}
\subsection{Experimental set-up and method}
\begin{figure}
    \centering
    \includegraphics[width=0.7\textwidth]{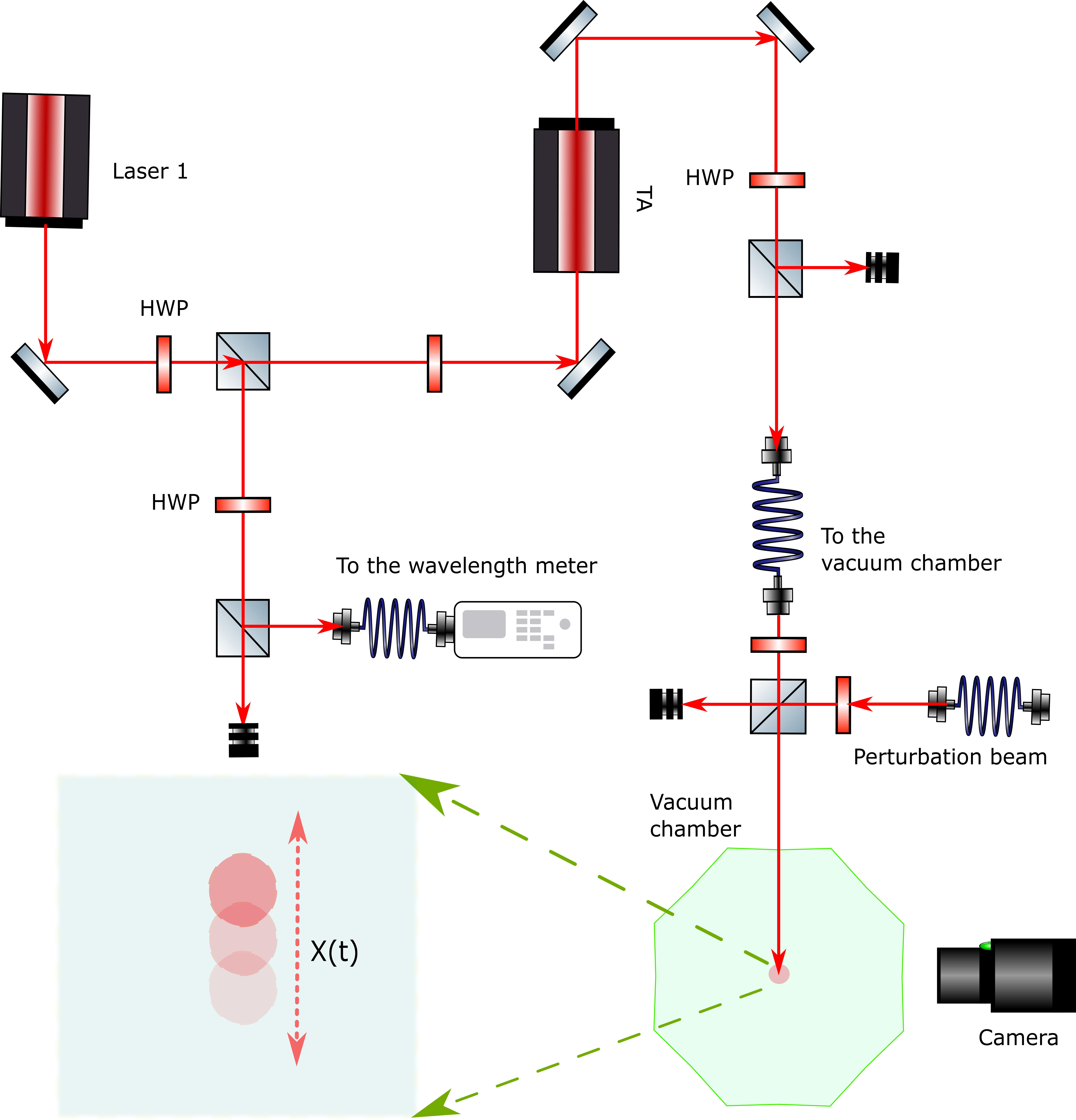}
    \includegraphics[width=0.7\textwidth]{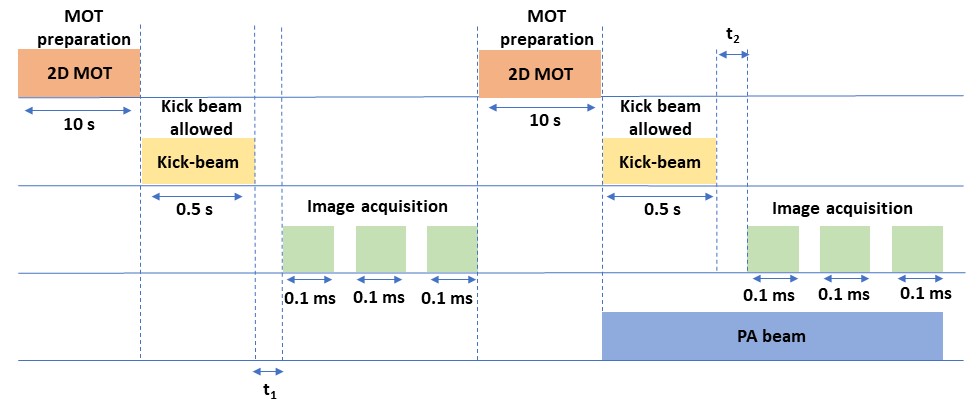}
    \caption{\textit{Upper Panel}: A schematic diagram of the experimental setup.  In the diagram, it can be observed that the driving beam and the PA-beam are merged and sent through the same path towards the trapped $^{39}$K cloud in the vaccuum chamber. \textit{Lower Panel}: Pulse sequence generated from LabVIEW used in this experiment. After the atoms are loaded in the MOT, the driving beam is switched on for a duration of 0.5 s. $t_{1}$ ($t_{2}$) is the time-delay between imparting the driving beam and measuring the response of the atoms without (with) the PA-beam on. Both $t_1$ and $t_2$ are varied from 0.2 ms to 30 ms (also called as $\tau_{postdrive}$) simultaneously.}
    \label{fig:Schematic diagram}
\end{figure}

A detailed description of our machine capable of trapping large numbers of $^{39}$K and $^{23}$Na atoms in a Magneto-Optical Trap (MOT) can be found in \cite{sutradhar2023fast}. This setup enables fast loading of atoms into the 3DMOT with the help of two independent 2D$^{+}$MOTs, resulting in a high signal-to-noise ratio fluorescence images for precise centroid positioning of the atomic clouds. In our experiment, as schematically depicted in Fig. \ref{fig:Schematic diagram} upper panel, we introduce two additional beams to the trapped cloud: a driving beam, red-detuned $6.5\Gamma$ from the $^{39}K$ D2 line, and a Photoassociation (PA) beam from a separate laser, tuned to a $^{39}K_2^*$ PA line at 390.976803 $THz$ \cite{Stwalley_Wang_1999}, with an accuracy of $\pm$1 MHz. This PA line, chosen for its relevance to single-photon excitation of $^{39}K_{2}^*$ just below the $4P_{3/2} + 4S_{1/2}$ level and being $\sim$ 1000 $\Gamma$ red-detuned, minimizes photon scattering by cold $^{39}$K atoms while facilitating measurements of strong molecular transitions.

\noindent The experiment is performed in two separate sequences and two different settings. Initially, we prepare the $^{39}$K atoms, which have been laser-cooled and trapped in a MOT. In our first setting, we allow only the driving beam onto the MOT, keeping the PA-beam blocked. While performing the second part of the experiment, we allow the driving beam and observe its effect on the cloud keeping the PA-beam ON throughout the experiment. In either of these settings, when we shine the driving beam with or without the PA-beam onto the trapped $^{39}$K atoms, we track the movement of the centroid of the cloud.

\noindent In order to give uniform optical force on the entire cross section of the $^{39}$K atom cloud, the diameter of the beam has been kept at 10 mm which is sufficient for a cloud of diameter $\sim 4$ mm. This beam shifts the center of the trapping region to align with the zero point of the optical force. We give a detailed timing sequence for measuring the  response function (Fig. \ref{fig:Schematic diagram} lower panel) of cold $^{39}$K atoms. Initially, cold atoms are loaded into the MOT for 10 s. Blocking the 2D$^{+}$MOT path, a driving beam is applied for 0.5 s, allowing the cold atomic cloud to reach a new equilibrium and to ensure that the dynamics of the cloud is solely influenced by the driving beam. After switching off the beam, we track the motion of the atoms via fluorescence imaging. \noindent A weak (0.05 $I_{sat}$) and off-resonant (5$\Gamma$ blue detuned) reference beam is directed at the MOT, followed by capturing 3 frames with an Andor ICCD camera. The reference beam prepared so as to not disturb the cold atoms, aids in position tagging. The sequence involves: 1) capturing the driven atomic cloud with the reference beam, 2) capturing without the atomic cloud but with the reference beam, and 3) capturing with both the reference beam and 3DMOT beams off to note background counts.
\par
\noindent In the next phase of our experiment, we first observed the motion of the atomic cloud under a driving beam (intensity 0.3 $I_{sat}$) alone. We then added a PA-beam to this setup, following the same procedure as the previous phase of our experiment. The motion of the cloud was recorded using the previously described method, ensuring the post-drive duration ($\tau_{post drive}$) matched with that of the driving beam-only observations. After capturing the images, we turned off the MOT. Within our selected range of parameters we observed the overall shift in the centroid position of the cold atomic cloud to be around 3 mm. This shift is more or less the same irrespective of the presence or absence of the PA-beam which has an intensity ($|E_{0}|^{2}$ per unit area) of 400 $mW/mm^{2}$.
\subsection{Experimental determination of the response function from the average displacement of the trapped cloud}
In this section we briefly outline the experimental method of determination of the response function from the measured average displacement discussed in ref. \cite{bhar2022measurements}.
The average displacement $\langle X(t)\rangle$ is related to the position response function $R(t)$ by the equation,
\begin{eqnarray}
 \langle X(t)\rangle &=&\int_{-\infty}^{t} R(t-t')f(t')dt'\label{xtRt}   
\end{eqnarray}
This gives the expectation value of the velocity on differentiation with respect to time $t$, $\langle \dot{X}(t)\rangle$ :
\begin{eqnarray}
\langle \dot{X}(t)\rangle &=&\int_{-\infty}^t \dot{R}(t-t')f(t')dt' \label{vtRt}
\end{eqnarray}
Here $f(t)$ is the external perturbing force, which in our experiment takes the form of a “top-hat function”:
\begin{eqnarray}
f(t) &=& \begin{cases}
  f_0, \text{for}\; -\infty<t<0\\
0, \text{for}\; t \leq -\infty, t \geq 0  
\end{cases}
\end{eqnarray}
Substituting $f(t)$ in Eq. (\ref{vtRt}), we get,
\begin{eqnarray}
    R(t)=-\frac{\langle \dot{X}(t)\rangle}{f_0}\label{RtfromVt}
\end{eqnarray}
\begin{figure}[H]
    \centering
    \includegraphics[width=.7 \textwidth]{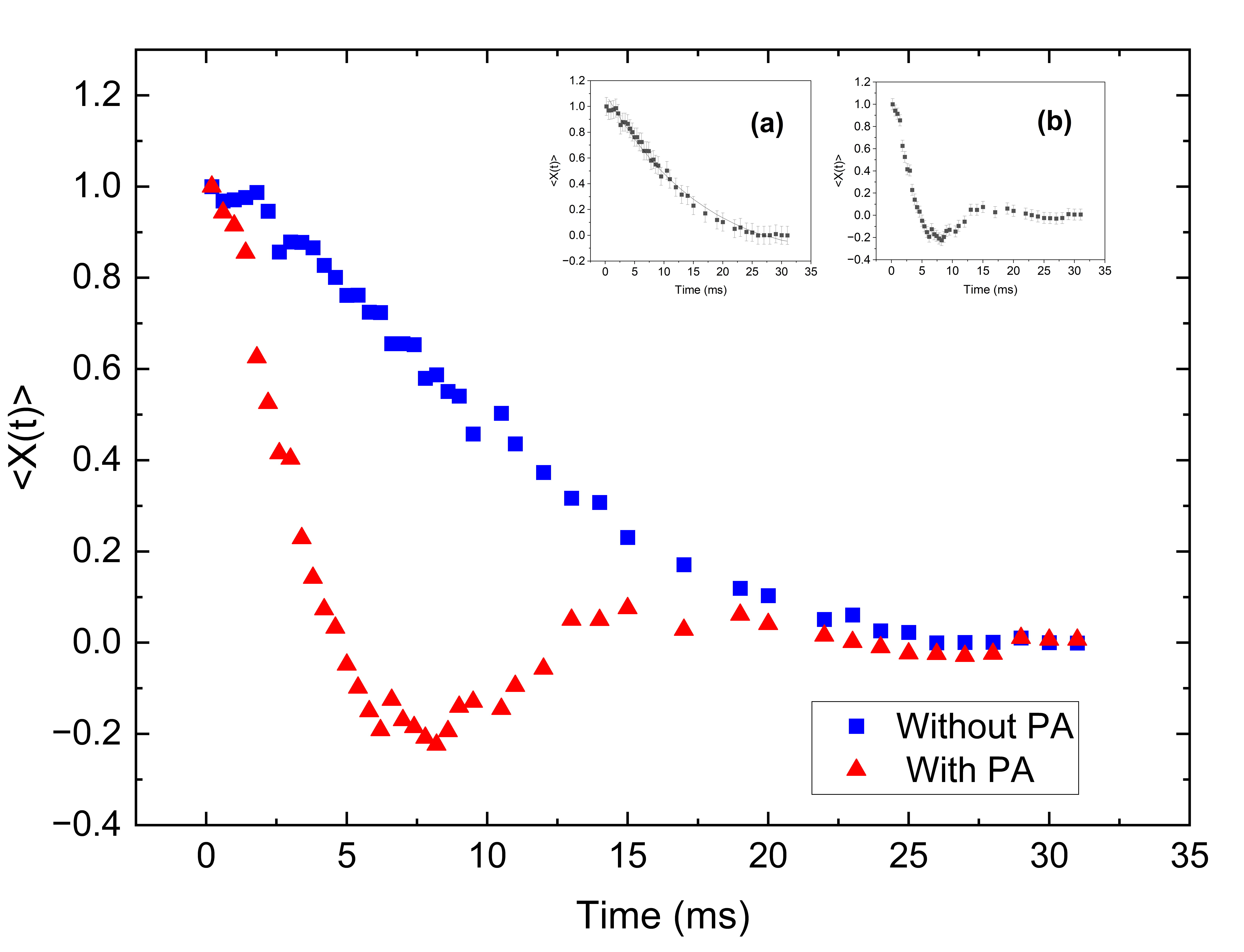}
    \caption{$\langle X(t)\rangle$ for both the incidents when only the driving beam is allowed to perturb the cold atomic cloud without and with the PA beam being present. [Inset] (a) The average displacement of cold cloud without and (b) in the presence of light-assisted interaction induced by the presence of Photoassociation laser. The error bars indicate $\pm 1\sigma$ statistical error estimated from several sets of data.}
    \label{fig:X_t}
\end{figure}
\noindent In the experiment we measure the average displacement $\langle X(t)\rangle$ of the cold atomic cloud as shown in Fig. \ref{fig:X_t}, and using the relation Eq. (\ref{RtfromVt}) we extract the response function $R(t)$. We observe in the inset (b) of Fig. \ref{fig:X_t} a collective oscillation of the cold cloud in the presence of the PA-beam while all other parameters are kept the same as in the inset (a) of Fig. \ref{fig:X_t}. Therefore we conclude that the collective oscillation arises due to inter-atomic interaction which is induced by the PA-beam. We record several sets of data to confirm the response of the cold atoms. The measurement statistical 1$\sigma$ errors are reported in the insets of Fig. \ref{fig:X_t} along with the data points.
\subsection{The role of the photoassociation laser frequency}
\begin{figure}[H]
    \centering
    \includegraphics[width = 0.7\textwidth]{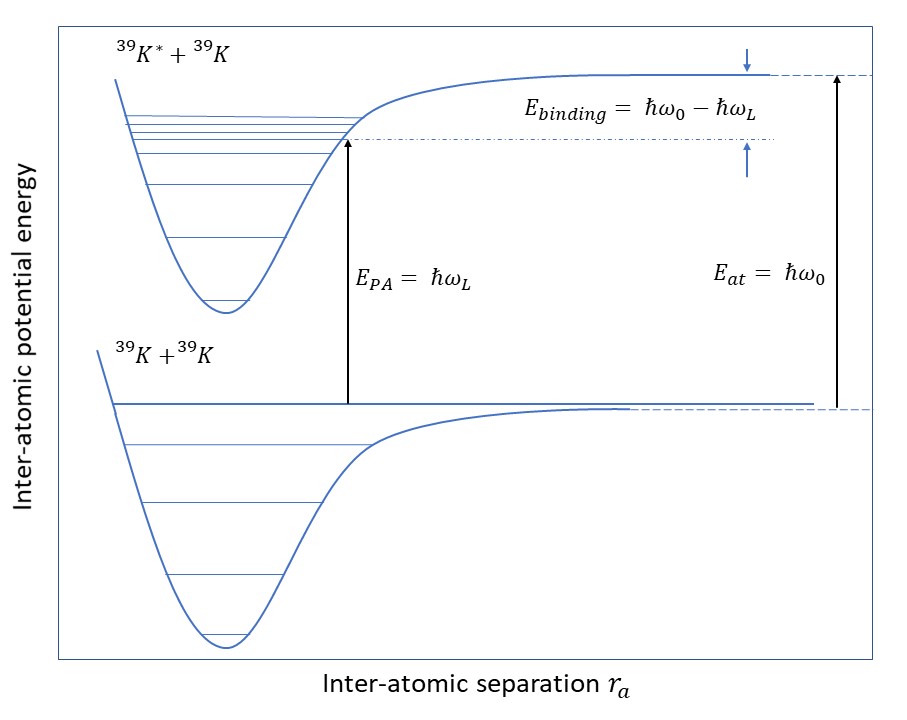}
    \caption{A schematic energy diagram showing an excited bound state formation using specific angular frequency ($\omega_L$) excitation laser. The upper (lower) potential energy curve shows the inter-atomic potential as a function of the inter-atomic separation of the excited (ground) molecular manifold.}
    \label{PA_potential}
\end{figure}

\noindent In Fig \ref{PA_potential} a schematic diagram of the photoassociation (PA) process is described \cite{jones2006ultracold}. There are discrete angular frequencies ($\omega_L$) at which the specific excited bound states can be addressed. Typically the linewidths of such excited molecular bound-states are in the range of $\sim$GHz. Near each of the molecular resonances the inter-atomic interaction strength also diverges. In our experiment, we choose  $\omega_L$ to be one such molecular resonance angular frequency. Generally in a PA spectroscopy experiment, the intensity of the PA laser is kept very high so as to induce significant atom loss from the trap. In contrast, we deliberately keep the PA laser intensity at a moderate level so that the atom loss is not significantly high while the inter-atomic interaction is still strong at resonance. Therefore, we can investigate the modification in the dynamics of the trapped cloud. In the theoretical analysis, discussed above, $\omega_L$ is taken as a parameter where the inter-atomic interaction is a maximum.

\section{Model of N-interacting particles in a thermal bath: Generalized Langevin equation and position response function}\label{theoretical_model}
\noindent This is a basic model for calculating the position response function for 
$N$ interacting particles, where $N$ is arbitrary. Using two interacting particles (i.e. $N=2$) in contact with a thermal bath as a toy model, we have calculated the response function of the centroid coordinate of the interacting particles under the influence of a laser field. 
\subsection{Description of the two-body model}
\noindent Starting from a Hamiltonian for a system of $N=2$ interacting particles coupled to a heat bath of independent harmonic oscillators and integrating out the bath degrees of freedom, we arrive at the quantum Langevin equation (QLE). We extract the position response function of the system from this QLE. This response function captures the collective behaviour observed in the experiment. 
\begin{figure}[H]
    \centering
    \includegraphics[width=.5\textwidth]{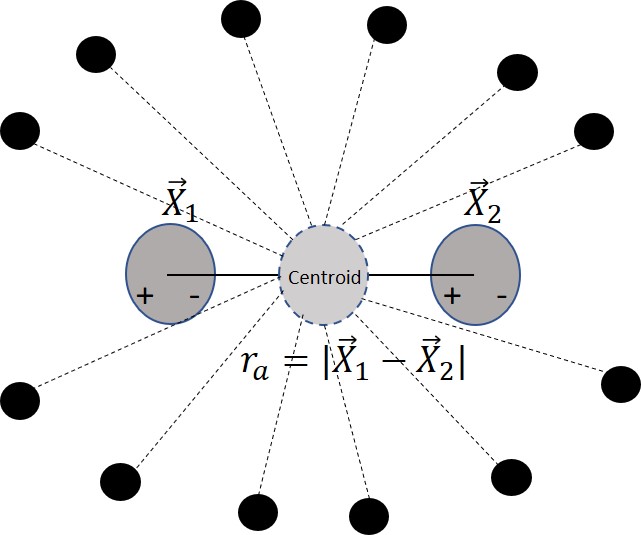}
    \caption{A schematic representation of the two particle system centroid or the centre of mass interacting with bath particles. The bath particles are harmonic oscillators and are represented by small particles and they do not interact with one another, hence the bath is called an IO bath. The system particles are represented by the two big gray solid circles, the centroid of which is shown by a dashed circle.}
    \label{twoparticle_schematic}
\end{figure}
To start with we consider the coupled system bath model, where the system is composed of two particles of the same mass ($M_1=M_2=1$) interacting with one another under the influence of an external laser field $E(t)$. The bath is composed of $K$ non interacting independent particles of mass $m_{\alpha}$ ($\alpha =1,2 ..., K$) and 
frequency $\omega_{\alpha}$ ($\alpha =1,2 ..., K$). This model is called the Independent-Oscillator (IO) model \cite{Ford1988}. There is a coupling of strength $c_{\alpha}$ ($\alpha =1,2 ..., K$) between the system coordinate (centroid coordinate $X_{c}=\sum_{i=1}^{2}X_i/2$) and bath coordinates. The Hamiltonian of the coupled system-bath model is given by, 
\begin{eqnarray}
    H &=&H_{s} +H_{b} +H_{c}\\
    H_s &=&\sum_{i=1}^{2}\left(\frac{P_i ^2}{2}+\frac{\Omega_0 ^2 X_i ^2}{2}+d_i E(t)\right)+\sum_{i,j,i\neq j}^{2} \frac{q^2}{2 r_a^3}X_i X_j\\
    H_b &=&\sum_{\alpha}\left(\frac{p_{\alpha}^2}{2m_{\alpha}}+\frac{m_{\alpha}\omega_{\alpha}^2x_{\alpha}^2}{2}\right)\\
    H_c&=&\sum_{\alpha}\left(\frac{c_{\alpha}^2}{2m_{\alpha}\omega_{\alpha}^2}X_{c}^2 -c_{\alpha}x_{\alpha}X_{c}\right)
\end{eqnarray}
Here, $H_s,H_b,H_c$ are respectively the system Hamiltonian, bath Hamiltonian and system-bath coupling Hamiltonian. $r_a=|X_1 -X_2|$ is the inter-atomic distance. A schematic diagram of the two particle centroid interacting with independent bath particles is shown in Fig. \ref{twoparticle_schematic}. $\Omega_0$ is the oscillator frequency of each of the system particles, 
$d_i, i=1,2$ are the dipole moments of the two system particles, each of charge $q$. 
The recipe for arriving at the QLE involves deriving the effective equations of motion of the system coordinates $X_i,P_i$ by integrating out the bath coordinates $x_{\alpha},p_{\alpha}$. Using Heisenberg equations of motion we get the QLE of the centre of mass coordinate of the interacting system : 
\begin{eqnarray}
    \ddot{X}_{c}+\Omega_0 ^2 X_c+\frac{q^2}{r_a^3} X_c-2\int_{t_0}^{t} ds \gamma(t-s)\dot{X}_c=\zeta(t)-q E(t)
\end{eqnarray}
where,
\begin{eqnarray}  
\gamma(t)&=&\sum_{\alpha}\frac{c_{\alpha}^2}{m_{\alpha}\omega_{\alpha}^2}\cos(\omega_{\alpha}t)\\
\zeta(t)&=&\sum_{\alpha}c_{\alpha}\left[ \left(x_{\alpha}(t_0)-\frac{c_\alpha}{m_\alpha \omega_\alpha ^2}X_c(t_0)\right)\cos(\omega_\alpha t)+\frac{p_\alpha(t_0)}{m_\alpha\omega_\alpha}\sin(\omega_\alpha t)\right]
\end{eqnarray}
are the dissipation kernel and noise terms, and $E(t)=E_0 \cos(\omega_L t)$, $|E_0|^2$ per unit area of the beam is the intensity of the laser and $\omega_L$ is the laser angular frequency. 
The pairwise interaction strength is a maximum at $\omega_L$, not at the resonant absorption angular frequency of photons by the atoms.
The bath information is only retained in the dissipation $\gamma(t)$ and noise $\zeta(t)$ terms. In the continuum limit, $K\longrightarrow \infty$ the thermal noise satisfies the following spectral properties,
\begin{eqnarray}
 \langle\zeta(t)\rangle&=&0\\
\langle\lbrace{\zeta(t),\zeta(t')\rbrace}\rangle&=&\frac{1}{2\pi}\int_{-\infty}^{\infty}d\omega Re[\gamma(\omega)] \hbar \omega \coth\left(\frac{\hbar \omega}{2k_B T}\right)e^{-i\omega t}\\
\langle[\zeta(t),\zeta(t')]\rangle&=&\frac{1}{\pi}\int_{-\infty}^{\infty}d\omega Re[\gamma(\omega)] \hbar \omega e^{-i \omega t}
\end{eqnarray}
The above relation is obtained considering that the bath is in thermal equilibrium at the initial time $t_0$ and is given by distribution function, $f(\omega)=1/(e^{(\hbar\omega/2k_B T)}-1)$. 
In the Fourier domain we can write the solution to the QLE,
\begin{eqnarray}
    X_c(\omega)&=&\frac{\zeta(\omega)-q E(\omega)}{-\omega^2+\omega_s ^2 -2i\omega\gamma(\omega)}= \frac{\zeta(\omega)-q E(\omega)}{(\omega- s_1) (\omega-s_2)} \label{Xcomega}
\end{eqnarray}
Here, $\omega_s^2=\Omega_0^2+\frac{q^2}{r_a^3}, s_1, s_2$ are the roots of the quadratic equation in the denominator of Eq. (\ref{Xcomega}). In order to extract the roots $s_1,s_2$ we need to model the thermal bath dissipation kernel $\gamma(t)$. Here we consider an Ohmic bath model \cite{satpathi2017quantum} for which $\gamma(t)=2\gamma_0 \delta(t)$ or $\gamma(\omega)=\gamma_0$. For the Ohmic bath, $s_1=-i\gamma_0-i\gamma_s, s_2=-i\gamma_0+i\gamma_s$, where $\gamma_s=\sqrt{\gamma_0^2-\omega_s^2}$. 
Using the fluctuation-dissipation theorem \cite{Kubo1966} which relates the Fourier transforms of the response function $R(t)$ and the position correlation function $C_X (t)=\langle\lbrace X_c(t),X_c(0)\rbrace\rangle/2$,
\begin{eqnarray}
    Im R(\omega)&=&\frac{1}{\hbar}\tanh{\left(\frac{\hbar\omega}{2k_B T}\right)}C_X(\omega)
\end{eqnarray}
we get $Im R(\omega)$ for two particles interacting with one another, coupled to an Ohmic bath in the presence of a laser field,
\begin{eqnarray}
    Im R(\omega)&=&\frac{\omega\gamma_0+q^2 E_0^2 \left[\delta(\omega-\omega_L)+\delta(\omega+\omega_L)\right]^2}{(\omega^2-s_1^2)(\omega^2-s_2^2)}
\end{eqnarray}
Using Kramer's Kronig relations we can calculate $Re R(\omega)$,
$$Re R(\omega)=\frac{1}{\pi}P\int_{-\infty}^{\infty}\frac{\omega'Im R(\omega')}{(\omega'^2-\omega^2)}d\omega'$$
and hence combining the real and imaginary parts we get the frequency dependent response function $R(\omega)=Re R(\omega)+i Im R (\omega)$, which on performing an inverse Fourier transform gives the time dependent response function $R(t)$, 
\begin{eqnarray}
    R(t)&=&\int_{-\infty}^{\infty} d\omega R(\omega)e^{-i\omega t}\\
    &=&\frac{e^{-\gamma_0 t}}{\gamma_s}\sinh{\gamma_s t}+Re\left(\frac{q^2 E_0^2 e^{-(\gamma_0-\gamma_s)t}s_2\left[\delta(s_1-\omega_L)+\delta(s_1+\omega_L)\right]^2}{4\gamma_0\gamma_s(\gamma_s^2-\gamma_0^2)}\right)\nonumber\\&&+Re\left(\frac{q^2 E_0^2 e^{-(\gamma_0+\gamma_s)t}s_1\left[\delta(s_2-\omega_L)+\delta(s_2+\omega_L)\right]^2}{4\gamma_0\gamma_s(\gamma_s^2-\gamma_0^2)}\right)
\end{eqnarray}
In a similar way one can solve the case of three interacting particles and find the response function. We have generalised the two particle system analysis to the context of a three particle system and found the QLE,
\begin{eqnarray}
    \ddot{X}_{c}+\Omega_0 ^2 X_c+\frac{2q^2}{r_a^3} X_c-3\int_{t_0}^{t} ds \gamma(t-s)\dot{X}_c=\zeta(t)-q E(t)
\end{eqnarray}
and the response function is given by,
\begin{eqnarray}
    R(t)&=&\frac{e^{-\gamma_0' t}}{\gamma_s'}\sinh{\gamma_s' t}+Re\left(\frac{q^2 E_0^2 e^{-(\gamma_0'-\gamma_s')t}s_2\left[\delta(s_1-\omega_L)+\delta(s_1+\omega_L)\right]^2}{4\gamma_0'\gamma_s'(\gamma_s'^2-\gamma_0'^2)}\right)\nonumber\\&&+Re\left(\frac{q^2 E_0^2 e^{-(\gamma_0'+\gamma_s')t}s_1\left[\delta(s_2-\omega_L)+\delta(s_2+\omega_L)\right]^2}{4\gamma_0'\gamma_s'(\gamma_s'^2-\gamma_0'^2)}\right)
\end{eqnarray}
Here $X_c=\sum_{i=1}^{3}X_i/3$, $\gamma_0'=3\gamma_0/2$, $\gamma_s'^2=\gamma_0'^2-\omega_s'^2$,$\omega_s'^2=\Omega_0^2+2q^2/r_a^3,s_1=-i\gamma_0'-i\gamma_s', s_2=-i\gamma_0'+i\gamma_s'$.
\subsection{$N$-particles pairwise interacting in a thermal bath}
We generalize the formalism further to a $N$ particle system. The QLE for a $N$ particle system reduces to,
\begin{eqnarray}
    \ddot{X}_{c}+\Omega_0 ^2 X_c+\frac{N q^2}{r_a^3} X_c-N\int_{t_0}^{t} ds \gamma(t-s)\dot{X}_c=\zeta(t)-q E(t)\label{QLEforNparticle}
\end{eqnarray}
and the response function is given by,
\begin{eqnarray}
    R(t)&=&\frac{e^{-\gamma_0' t}}{\gamma_s'}\sinh{\gamma_s' t}+Re\left(\frac{q^2 E_0^2 e^{-(\gamma_0'-\gamma_s')t}s_2\left[\delta(s_1-\omega_L)+\delta(s_1+\omega_L)\right]^2}{4\gamma_0'\gamma_s'(\gamma_s'^2-\gamma_0'^2)}\right)\nonumber\\&&+Re\left(\frac{q^2 E_0^2 e^{-(\gamma_0'+\gamma_s')t}s_1\left[\delta(s_2-\omega_L)+\delta(s_2+\omega_L)\right]^2}{4\gamma_0'\gamma_s'(\gamma_s'^2-\gamma_0'^2)}\right)\\
    &=&\frac{e^{-\gamma_0' t}}{\gamma_s'}\sinh{\gamma_s' t}+R_2(t)+R_3(t) \label{RTNparticle}
\end{eqnarray}
where,
\begin{eqnarray}
    R_2(t)=\begin{cases}
        Re\left(\frac{q^2 E_0^2 e^{-(\gamma_0'-\gamma_s')t}s_2}{4\gamma_0'\gamma_s'(\gamma_s'^2-\gamma_0'^2)}\right), \text{for}\; \omega_L=|s_1|\\
        0
    \end{cases}\\
        R_3(t)=\begin{cases}
        Re\left(\frac{q^2 E_0^2 e^{-(\gamma_0'+\gamma_s')t}s_1}{4\gamma_0'\gamma_s'(\gamma_s'^2-\gamma_0'^2)}\right), \text{for}\; \omega_L=|s_2|\\
        0
        \end{cases}
\end{eqnarray}
Here $X_c=\sum_{i=1}^{N}X_i/N$, $\gamma_0'=N\gamma_0/2$, $\gamma_s'^2=\gamma_0'^2-\omega_s'^2$,$\omega_s'^2=\Omega_0^2+ N q^2/R^3,s_1=-i\gamma_0'-i\gamma_s', s_2=-i\gamma_0'+i\gamma_s'$. Note that the dissipation kernel and the noise terms remain the same as in the two particle case, the only change appears in $X_c$. In the absence of interactions when $\gamma_0'>>\Omega_0$, we get the expected overdamped behaviour in the position response function, however if we include interactions keeping $\gamma_0', \Omega_0$ unchanged we notice an oscillatory behaviour in the position response function. This is rather surprising but it can be understood from  a comparison of the two time scales $\gamma_0'^{-1}$ and $\gamma_s'^{-1}$. 
In the absence of interactions, $\gamma_0'^2>\gamma_s'^2$ and both are real, resulting in an exponentially damped behaviour of the position response function. Inclusion of interactions leads to the condition, $\gamma_0'^2<\gamma_s'^2$ where $\gamma_s'$ is imaginary, resulting in 
an oscillatory behaviour of the position response function. 
The response function given by Eq. (\ref{RTNparticle}) captures the transition from damped to oscillatory behaviour without and with interactions, respectively. 
We have used Eq. (\ref{RTNparticle}) in the paper for comparing our theory with the experimentally observed data.

\section{Experimental Results and Comparison with theory}
 \begin{figure}
    \centering
    \includegraphics[width=0.8 \textwidth]{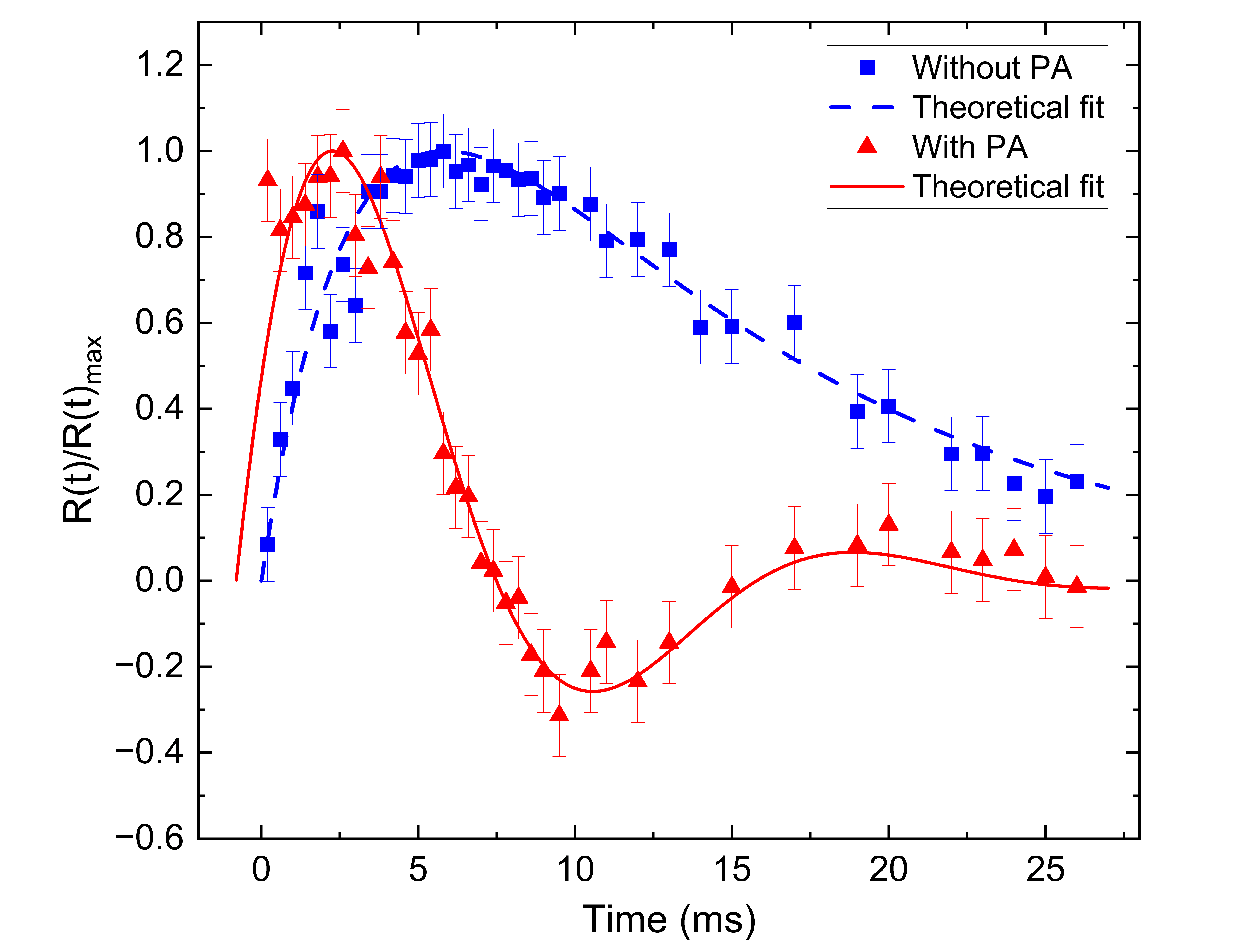}
    \caption{The normalized response function for both the incidents with only the driving beam perturbing the cold atomic cloud in the overdamped regime without and with the PA beam present. }
    \label{fig:NMS_sept-16_w_kick}
\end{figure}
Following the methods described in section \ref{section2} we measure the centroid position of the cloud as a function of time and extract the position response function, $R(t)$. We normalize R(t) w.r.t. its maximum value. We plot the normalized position response function of the cold atomic cloud as a function of $\tau_{postdrive}$. The resulting observation is summarized in Fig \ref{fig:NMS_sept-16_w_kick}. The transition from an overdamped motion to an underdamped oscillatory motion is clearly visible in the same run of an experiment upon the addition of an inter-atomic interaction induced by the PA-beam.
The derivation of the Generalized Quantum Langevin equation (QLE) and the extraction of the position response function of the centroid of $N-$ interacting atoms under the influence of an external laser are outlined in the section \ref{theoretical_model}.  The QLE and the position response functions are given respectively by the equations:
\begin{align}
    \ddot{X}_{c}+\Omega_0 ^2 X_c &+\frac{N q^2}{r_a^3} X_c-N\int_{t_0}^{t} ds \gamma(t-s)\dot{X}_c=\zeta(t)-q E(t)\label{QLEforNparticle}\\
    R(t) &=\frac{e^{-\gamma_0' t}}{\gamma_s'}\sinh{\gamma_s' t}+R_2(t)+R_3(t)\label{RTNparticle}\\
    R_2(t)&=\begin{cases}
        Re\left(\frac{q^2 E_0^2 e^{-(\gamma_0'-\gamma_s')t}s_2}{4\gamma_0'\gamma_s'(\gamma_s'^2-\gamma_0'^2)}\right), \text{for}\; \omega_L=|s_1|\\
        0
    \end{cases}\\
        R_3(t)&=\begin{cases}
        Re\left(\frac{q^2 E_0^2 e^{-(\gamma_0'+\gamma_s')t}s_1}{4\gamma_0'\gamma_s'(\gamma_s'^2-\gamma_0'^2)}\right), \text{for}\; \omega_L=|s_2|\\
        0
        \end{cases}
\end{align}

\noindent Here $X_c=\sum_{i=1}^{N}X_i/N$, is the centroid displacement, $\gamma_0'=N\gamma_0/2$, $\gamma_s'^2=\gamma_0'^2-\omega_s'^2$,$\omega_s'^2=\Omega_0^2+ N q^2/r_a^3,s_1=-i\gamma_0'-i\gamma_s', s_2=-i\gamma_0'+i\gamma_s'$ and $E(t)=E_0 cos(\omega_L t)$ is the external laser field. $\omega_L$ is the angular frequency of the photoassociation laser field which induces the interaction between the atoms. In the experiment, we measure the average displacement $\langle X_c(t)\rangle$ which is related to the position response function $R(t)$ as, $R(t)=-\langle \dot{X_c}(t)\rangle/f_0$ \ref{section2}. The strength of the interaction depends on $\omega_L$ following the details of the two-body inter-atomic potential \cite{jones2006ultracold, ridinger2011photoassociative}; however, in our description, we have considered $\omega_L$ to be a parameter. 

$\gamma_0'$ pertains to the damping coefficient due to the optical molasses and the frequency constant $\Omega_0$ stems from the magnetic field gradient of the MOT. Three distinct cases appear for the motion of the atomic cloud in the absence of inter-atomic interactions, depending on the values of $\gamma_0', \Omega_0$: an overdamped motion for $\gamma_0'>\Omega_0$, a critically damped motion for $\gamma_0'=\Omega_0$ and an underdamped motion for $\gamma_0'<\Omega_0$.Since we can choose a suitable parameter space in our experiment to operate either in the overdamped or underdamped regime, we consider the overdamped case. We have fitted the data using Eq.(\ref{RTNparticle}) which is the position response function for $N$ interacting atoms under the influence of an external laser, stemming from a generalised QLE (Eq.(\ref{QLEforNparticle})). 

\begin{figure}[H]
    \centering
    \includegraphics[width= 0.7\textwidth]{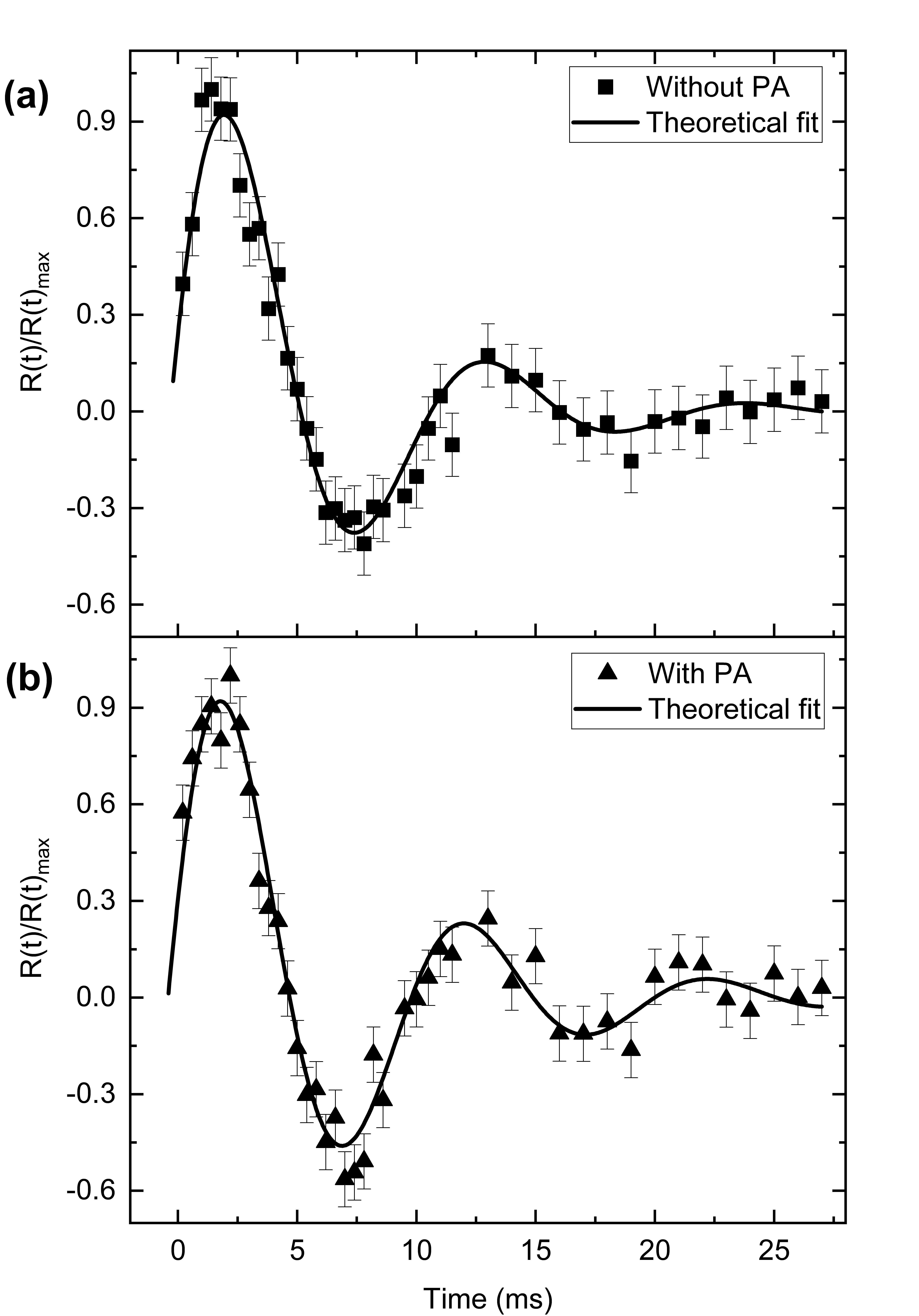}
    \caption{The molasses parameters are chosen such that the normalized position response function of the cold atomic cloud is in the underdamped regime. In \textbf{(a)} The data has been recorded in the absence of the 'PA-beam' and \textbf{(b)} in the presence of the 'PA-beam'. The data has been fitted here with the equation of the position response function which gives $\omega_{s}' = 2\pi\times90$ Hz for (a) and $\omega_{s}' = 2\pi\times102$ Hz for (b).}
    \label{fig:NMS_sept-16_w_wo_PA}
\end{figure}

\noindent We extract a damping coefficient, $\gamma_0'$, using the fit to the data. This damping coefficient is $\sim 1.23\times 10^{-23}$ kg/sec. In the absence of interactions when $\gamma_0'>>\Omega_0$, we get the expected overdamped behaviour in the position response function.

\begin{figure}[H]
    \centering
    \includegraphics[width=0.8\textwidth]{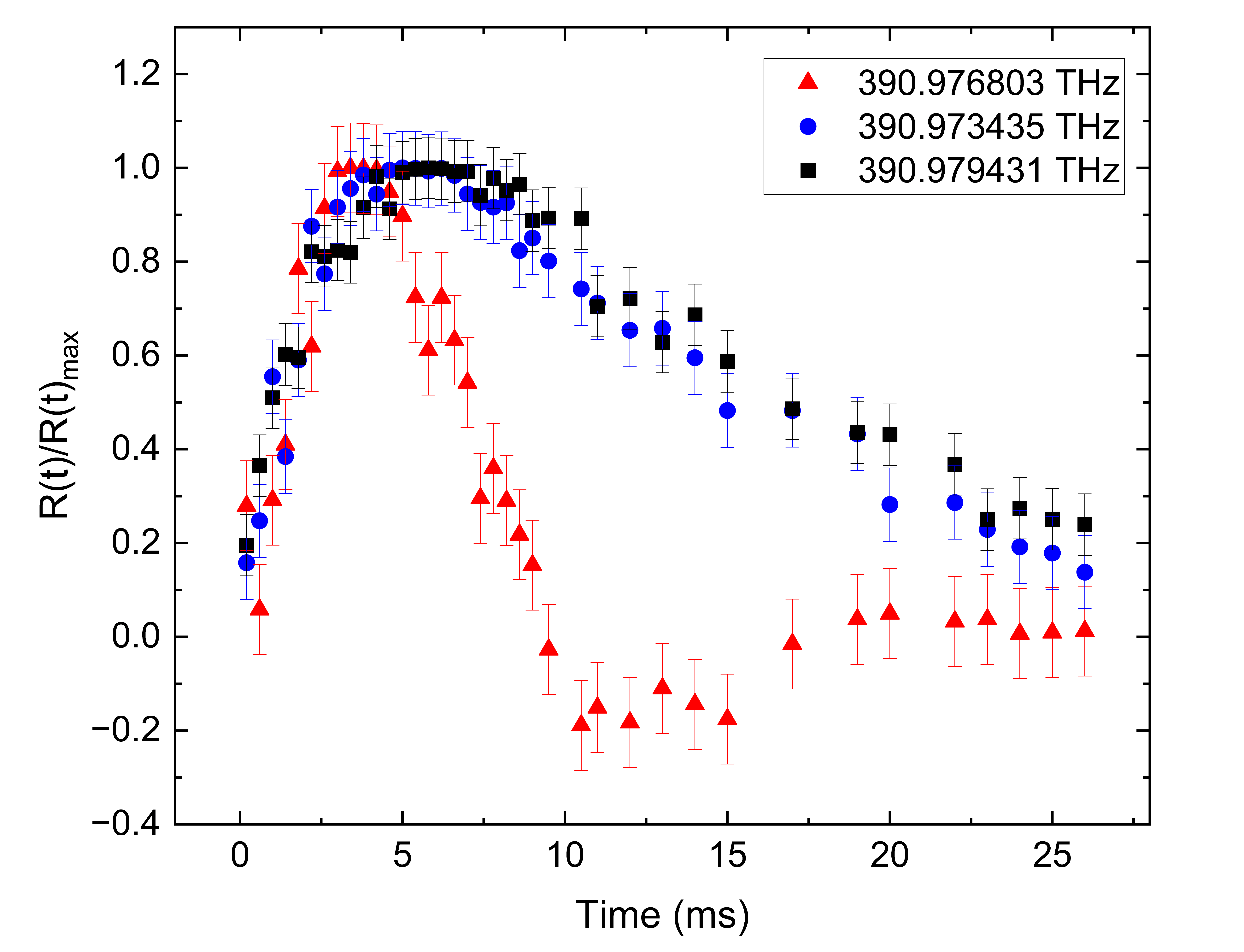}
    \caption{The normalized response function with the PA-beam kept ON at a fixed intensity while its frequency is changed, once above and once below a particular molecular transition frequency. Only when the laser is kept at the molecular transition line ($\omega_L= 2\pi\times$ 390.976803 $THz$), the oscillatory behaviour can be observed whereas if the PA-laser frequency is tuned to below or above the molecular transition line (which are 390.973435 $THz$ and 390.979431 $THz$ in the figure), the clear oscillatory nature is absent.}
    \label{@diff_wavenumber}
\end{figure}

 \noindent However, an oscillatory behaviour emerges in the position response function when we include atom-atom interactions keeping $\gamma_0', \Omega_0$ unchanged. This is rather surprising but it can be understood from a comparison of the two time-scales $\gamma_0'^{-1}$ and $\gamma_s'^{-1}$. In the absence of interactions, $\gamma_0'^2>\gamma_s'^2$ and both are real, resulting in an exponentially damped behaviour of the position response function Eq.(\ref{RTNparticle}). Inclusion of interactions leads to the condition, $\gamma_0'^2<\gamma_s'^2$ where $\gamma_s'$ is imaginary, resulting in an oscillatory behaviour of the position response function Eq.(\ref{RTNparticle}). 

\noindent We investigated the effect of inter-atomic interaction on the position response function in the under-damped regime as well. In Fig \ref{fig:NMS_sept-16_w_wo_PA}, we summarize our observation in this regime. In Fig \ref{fig:NMS_sept-16_w_wo_PA}(a) the normalised position response function of the cold cloud (as a function of time) is shown without the light-assisted atom-atom interaction, whereas in Fig \ref{fig:NMS_sept-16_w_wo_PA}(b) the atom-atom interaction has led to a marked increase in the oscillation frequency by more than 10$\%$. The response function given by Eq. (\ref{RTNparticle}) captures this enhancement of the frequency of oscillations in the under-damped regime without and with interactions, respectively.

Figs. \ref{fig:NMS_sept-16_w_kick} and \ref{fig:NMS_sept-16_w_wo_PA} show a comparison between the experimentally measured response function and the theoretical response function from Eq.(\ref{RTNparticle}) for the overdamped and underdamped regimes respectively. The theoretical expression shows a very good fit to the experimental data for both cases. The dipole moment of a $^{39}$K atom in its ground state is much smaller than that of a Rydberg atom \cite{2015_Browaeys}, or an atom possessing high dipole moment such as Chromium \cite{Pfau2005} at sub-micro Kelvin temperatures. Therefore, the angular dependence of the dipole-dipole interaction has been neglected in our theoretical modelling. 

We conducted additional experiments where the PA-beam was kept ON at the same intensity but its frequency was shifted away from the molecular transition frequency. In Fig. \ref{@diff_wavenumber}, we show the response function for three different values of $\omega_L$. Dramatically, the oscillatory behaviour is seen only when $\omega_L$ is tuned to a molecular resonance whereas shifting $\omega_L$ slightly ($\pm 3$ GHz) takes it to an overdamped motion. This resonance tracking is beyond the discussed theoretical model (Eq.(\ref{RTNparticle})) because $\omega_L$ is taken as a fixed parameter. The atom-atom interaction strength as a function of $\omega_L$ is discussed in the section \ref{theoretical_model}.
\par

In our magneto-optically trapped cold atomic cloud of $^{39}K$, the inter-atomic interactions are described by the s-wave scattering length, $a$ (for temperatures $\leq$ 2 mK). For large PA beam detuning from the atomic transition ($|\delta| \gg \Gamma$, where $\Gamma$ is the decay rate of the atomic transition), the scattering length $a(\omega_L)$ is adjusted by the PA laser frequency $\omega_L$ and the PA-beam intensity $I$ \cite{2015_Julienne, 2011_Ye} as $ a(\omega_L)=a_{bg}+ l_{opt}\frac{\Gamma_m}{\omega_{L}-(\omega_n + s_n I)}$. Here, $a_{bg}$ is the background scattering length, $\Gamma_m$ is the spontaneous emission rate from the excited molecular state, and $\omega_n + s_n I$ is intensity-dependent shift of the resonance frequency. The optical length $l_{opt}$, indicative of the PA line-strength, incorporates the influence of the PA laser. With the background scattering length of $^{39}K$ atoms being negative ($-a_{bg}\simeq 33a_0$), the inherent inter-atomic interaction is attractive and corresponds to an overdamped behavior of the position response function without the PA beam. By applying the PA beam near a molecular resonance the interaction alters, enabling a shift from an overdamped to an underdamped behaviour of the position response function.
 \par
\section{Conclusion and outlook}
\noindent In this study, we alter the inter-atomic interactions by exposing a cold cloud of $^{39}K$ atoms to a beam tuned to one of the PA lines. Tracking the position response function, we offer a novel approach to detect resonant enhancement of interaction strength in cold atoms. 
 Extending beyond existing theoretical analysis \cite{bhar2022measurements}, our study incorporates light-assisted inter-atomic interactions into a generalized Langevin formalism, focusing on their impact on the position response function over a significant range of time scales, while disregarding molecule formation due to its brevity in this context. 
Our theoretical analysis aligns well with our experimental observations, setting the stage for more intricate future models of inter-atomic interactions as well as by including angle dependent dipole-dipole interactions in quantum-degenerate clouds. Moreover, our experimental system also allows us to study a mass-imbalanced cold atomic mixture of $^{23}$Na and $^{39}$K which opens avenues to explore the position response function amidst varying intra- and inter-species atomic interactions.

\section{Acknowledgement}
\noindent The authors acknowledge the funding from Department of science and Technology, India. The authors also thank S. Majumder and G. pal for discussions as well as helping with the instruments. The support from the mechanical and electrical workshops in Raman Research Institute is appreciated for conducting the experiments successfully.

\bibliography{Response_Interaction}

\end{document}